\documentclass[12pt,a4paper]{article}
\usepackage{amsfonts,latexsym}
\usepackage{amsmath,amssymb}
\usepackage{graphicx,color}
\usepackage{multirow}
\usepackage{slashbox}
\newcommand{\nn}{\nonumber}

\begin{document}
\vspace{12mm}

\begin{center}
{{{\Large {\bf Onset of rotating scalarized black holes in Einstein-Chern-Simons-Scalar theory}}}}\\[10mm]

{Yun Soo Myung$^a$\footnote{e-mail address: ysmyung@inje.ac.kr} and De-Cheng Zou$^{a,b}$\footnote{e-mail address: dczou@yzu.edu.cn}}\\[8mm]

{${}^a$Institute of Basic Sciences and Department  of Computer Simulation, \\ Inje University, Gimhae 50834, Korea\\[0pt] }

{${}^b$Center for Gravitation and Cosmology and College of Physical Science and Technology, Yangzhou University, Yangzhou 225009, China\\[0pt]}
\end{center}
\vspace{2mm}

\begin{abstract}
We perform the stability analysis of the Kerr  black hole
in the  Einstein-Chern-Simons-Scalar theory with quadratic scalar coupling.   For positive coupling parameter ($\alpha>0$), we introduce the (2+1)-dimensional  hyperboloidal  foliation method to show that
the Kerr black hole is unstable against  scalar-mode
perturbation.  In  case of $\alpha<0$, it is shown that  the Kerr black hole is unstable without any $\alpha$-independent bound on the rotation parameter  $a$ when using the the hyperboloidal  foliation method.
The disappearance of the $a$-bound  is  confirmed analytically
by  implementing the (1+1)-dimensional  linearized scalar equation.
\end{abstract}

\newpage
\renewcommand{\thefootnote}{\arabic{footnote}}
\setcounter{footnote}{0}


\section{Introduction}

Spontaneous scalarization is regarded as an important mechanism to obtain black holes with scalar hair numerically. It was  realized  in the models where
a scalar field   couples either to  Gauss-Bonnet term~\cite{Doneva:2017bvd,Silva:2017uqg,Antoniou:2017acq} or to Maxwell term~\cite{Herdeiro:2018wub} with a coupling function. The appropriate coupling function $f(\phi)$ includes quadratic, exponential scalar functions, and others.

Importantly, the linearly tachyonic  instability of  static  Schwarzschild~\cite{Myung:2018iyq} or Reissner-Nordstr\"{o}m  black hole~\cite{Myung:2018vug}  are considered  as onset of spontaneous scalarization.
As the instability develops and the scalar grows, nonlinear terms become important increasingly and quench   the instability eventually.
Accordingly, these nonlinear terms control the endpoint and properties of infinite $n=0,1,2,\cdots$ scalarized black holes.
It turns out that the $n=0$ scalarized  black hole is stable,
while all excited $n=1,2,3,\cdots$ scalarized black holes are unstable~\cite{Zou:2020zxq}.

However,  the  parity-odd Chern-Simons term is hard to activate these static black holes because
these are parity-even solutions~\cite{Ayzenberg:2013wua,Myung:2019wvb}. The Kerr is a parity-odd solution, implying that it may acquire modifications from the Chern-Simons term.  One notes that Schwarzschild black hole was stable in the Einstein-Chern-Simons gravity with linear scalar coupling~\cite{Molina:2010fb,Moon:2011fw,Kimura:2018nxk,Macedo:2018txb}. It  shows  the reason to ``why static  scalarized  black holes do not appear from the linear scalar coupling to the Chern-Simons term".

On the other hand, the spontaneous scalarization of rotating black holes in the Einstein-Gauss-Bonnet-Scalar (EGBS) theory with positive coupling parameter $\lambda^2$ has been studied in~\cite{Cunha:2019dwb,Collodel:2019kkx}. Here,  the rotation parameter $a$ suppresses scalarization and this case includes the scalarization of Schwarzschild black hole ($M/\lambda \le 0.587$) as the non-rotating limit of $a\to 0$.    Recently, the tachyonic instability condition  of Kerr black holes is found as an $\eta$-independent  bound of $a/M \ge 0.5$ in the EGBS theory with  negative coupling parameter $\eta$ when  solving the (1+1)-dimensional linearized scalar  equation~\cite{Dima:2020yac}. It was confirmed analytically in~\cite{Hod:2020jjy} and numerically in~\cite{Doneva:2020nbb}. This means that the spin-induced scalarized  black holes were constructed in the EGBS theory with negative coupling parameter~\cite{Herdeiro:2020wei,Berti:2020kgk}. Also, there exists a minimum rotation of $a_{\rm min}=0.5$ and below which no instability could be triggered, irrespective of any values of coupling parameter~\cite{Zhang:2020pko}.

In this work, we investigate the onset of rotating scalarized black holes in the Einstein-Chern-Simons-Scalar (ECSS) theory with quadratic scalar coupling.
For positive coupling parameter $\alpha$, it is meaningful to mention  that the Kerr black hole was unstable under the scalar mode perturbation~\cite{Gao:2018acg}.
Here, we wish to  study the onset of rotating scalarized black holes in ECSS theory with negative  coupling parameter.
First of all, we will introduce a (2+1)-dimensional  hyperboloidal foliation method to show  the tachyonic instability of Kerr black hole.
Further, an analytic approach  will be introduced to investigate  whether  there is  the $\alpha$-independent  bound on $a$ or not  in the ECSS theory with negative coupling parameter.

\section{ECSS theory}

We start with the ECSS theory  given by
\begin{eqnarray}
S_{\rm ECSS}=\frac{1}{16 \pi}\int d^4 x\sqrt{-g} \Big[
R-\frac{1}{2}(\partial \phi)^2+\alpha \phi^2{}^{*}RR \Big]\label{Action}
\end{eqnarray}
where ${}^{*}RR$ is  the
Chern-Simons term
\begin{equation}
{}^{*}RR={}^{*}R^{\eta~\mu\nu}_{~\xi}R^\xi_{~\eta\mu\nu}
\end{equation}
with  the dual Ricci tensor
\begin{eqnarray}
{}^{*}R^{\eta~\mu\nu}_{~\xi}=\frac{1}{2}\epsilon^{\mu\nu\rho\sigma}R^{\eta}_{~\xi\rho\sigma}.
\end{eqnarray}
Although  ${}^{*}RR$ is a topological term  in four dimensions, but
it  becomes dynamical due to  coupling with  the scalar.  The Chern-Simons term
does not activate a scalar monopole field in the static black hole spacetime, but it will activate  scalar modes in the Kerr black hole spacetime.
The coupling parameter $\alpha$ is now allowed to have  positive ($\alpha>0$) and negative ($\alpha<0$) except $\alpha=0$.

The variations of action (\ref{Action}) with respect to  $g_{\mu\nu}$ and $\phi$ lead to the
two equations as
\begin{eqnarray}
&&G_{\mu\nu}=\frac{1}{2}\partial_{\mu}\phi\partial_{\nu}\phi
- \frac{1}{4}g_{\mu\nu}(\partial \phi)^2 -4\alpha C_{\mu\nu},\label{eq-1}\\
&&\nabla^2\phi=-2\alpha~ {}^{*}RR \phi, \label{eq-2}
\end{eqnarray}
where the Cotton tensor  $C_{\mu\nu}$ is given by
\begin{eqnarray}\label{cotton}
C_{\mu\nu}=\nabla_{\rho}(\phi^2)~\epsilon^{\rho\sigma
\gamma}_{~~~~(\mu}\nabla_{\gamma}R_{\nu)\sigma}+\frac{1}{2}\nabla_{\rho}\nabla_{\sigma}
(\phi^2)~\epsilon_{(\nu}^{~~\rho \gamma
\delta}R^{\sigma}_{~~\mu)\gamma \delta}.
\end{eqnarray}

Selecting  the no-scalar hair (by denoting  the ``overbar'')
 \begin{equation}
\bar{\phi}=0,
\end{equation}
Eqs.(\ref{eq-1})-(\ref{eq-2}) imply  the axisymmetric Kerr spacetime
\begin{eqnarray}
ds_{\rm Kerr}^2&=&\bar{g}_{\mu\nu}dx^{\mu}dx^{\nu}\nn\\
&=& -\frac{\Delta}{\rho^2}(dt-a \sin^2 \theta d\varphi)^2+\frac{\rho^2}{\Delta}dr^2 \nn\\
&+&\rho^2 d\theta^2 +\frac{\sin^2 \theta}{\rho^2}[a dt -(r^2+a^2)d\varphi^2]   \label{kerr-sol}
\end{eqnarray}
with mass $M$, angular momentum per mass $a=J/M>0$, $\Delta=r^2-2Mr +a^2$, and $\rho^2=r^2+a^2 \cos^2\theta$.
The outer and inner horizons are determined by imposing $\Delta=0$ as
\begin{equation}
r_\pm=M[1\pm \sqrt{1-a^2/M^2}].
\end{equation}
In this case, the Chern-Simons term takes the non-vanishing background quantity as
\begin{equation} \label{CS}
{}^{*}\bar{R}\bar{R}=\frac{96aM^2r\cos \theta(3r^2-a^2\cos^2\theta)(r^2-3a^2\cos^2\theta)}{(r^2+a^2\cos^2\theta)^6},
\end{equation}
whose limit of $a\to 0$ is zero. This implies that there is no contribution from static black holes.

\section{Onset of rotating scalarization}

Now,  let us  introduce the perturbations
around the background  as
\begin{eqnarray} \label{m-p}
g_{\mu\nu}=\bar{g}_{\mu\nu}+h_{\mu\nu},~~\phi=0+\delta\phi.
\end{eqnarray}
The linearized equation to (\ref{eq-1}) can be written by
\begin{eqnarray}\label{pertg}
\delta R_{\mu\nu}(h)=0
\end{eqnarray}
with
\begin{eqnarray}\label{cottonp0}
\delta
R_{\mu\nu}(h)&=&\frac{1}{2}\left(\bar{\nabla}^{\gamma}\bar{\nabla}_{\mu}
h_{\nu\gamma}+\bar{\nabla}^{\gamma}\bar{\nabla}_{\nu}
h_{\mu\gamma}-\bar{\nabla}^2h_{\mu\nu}-\bar{\nabla}_{\mu} \bar{\nabla}_{\nu} h\right).\label{cottonp1}
\end{eqnarray}
The linearized scalar equation around the Kerr black hole leads to
\begin{eqnarray}
 \Big(\bar{\nabla}^2-\mu^2_{\rm eff}\Big)\delta\phi=0,\label{phi-eq2}
\end{eqnarray}
where the effective mass is given by
\begin{equation}
\mu^2_{\rm eff}=-2\alpha {}^{*}\bar{R}\bar{R}.
\end{equation}
Actually, a stability analysis for the Kerr black hole with (\ref{pertg}) is the same as in the general relativity, implying that there are  no exponentially growing tensor modes. Therefore, the stability of Kerr black hole will be determined solely by the linearized scalar equation (\ref{phi-eq2}) in the ECSS theory.
Observing (\ref{CS}) roughly,  its sign  would be positive or  negative.
From now on, we consider two cases of $\alpha>0$ and $\alpha<0$.

\subsection{$\alpha>0$ case}
\begin{figure*}[t!]
   \centering
  \includegraphics{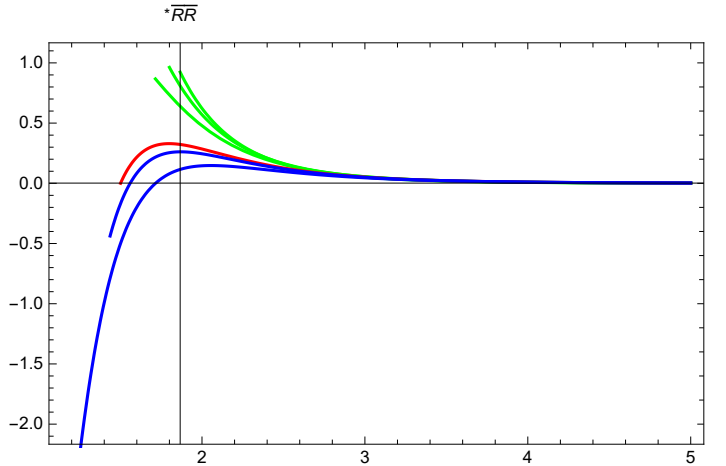}
\caption{Graphs for ${}^{*}\bar{R}\bar{R}$ with $M=1$ and $\theta=0$ as function of $r\in[r_+,5]$ and $a$. From top to bottom, we denote each graph for
 $a=0.5(r_+=1.866),0,6(r_+=1.8),0.7(r_+=1.714),0.866(r_+=1.5),0.9(r_+=1.436),0.99(r_+=1.141)$. Surely, $a=0.866$ (red curve) is the boundary between positive and negative values near the outer horizon.}
\end{figure*}

First of all, we examine the ${}^{*}\bar{R}\bar{R}$-term in (\ref{CS}) from Fig.1.
For $\theta=0$ and $a/M^2\le 0.866$ with $M=1$, ${}^{*}\bar{R}\bar{R}$ remains positive definite and monotonic  in $r$.
Choosing $\theta=0$ is closely related to finding the instability bound for $l\to\infty$-scalar mode in the ECSS theory with $\alpha<0$.
For $a/M^2> 0.866$, ${}^{*}\bar{R}\bar{R}$ becomes negative and thus, the high rotation can make the effective mass ($\mu^2_{\rm eff}=-2\alpha {}^{*}\bar{R}\bar{R}$) less negative or even positive near the horizon for $\alpha>0$. It suppresses the effect of scalarization. However, if one considers the EGBS theory, this suppressing bound is given by $a/M^2> 0.5$~\cite{Cunha:2019dwb}. Considering the opposite case of $\alpha<0$, this may be the enhancing (instability) bound [$(a/M)>(a/M)_{\rm crit}=0.5$] for the  scalarization in the EGBS theory with negative coupling parameter. Similarly, $(a/M)>(a/M)_{\rm crit}=0.866$ may be the enhanced (instability) bound for  scalarization in the ECSS theory with negative coupling parameter.
\begin{figure*}[t!]
   \centering
  \includegraphics{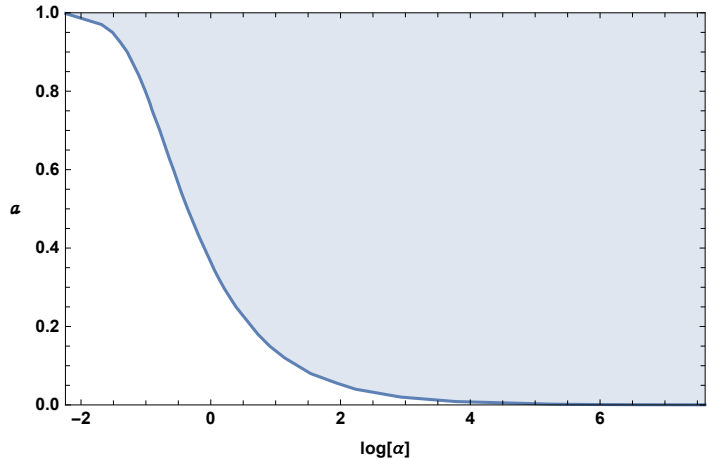}
\caption{$\alpha$-dependent instability bound positive $\alpha$. Shaded region denotes unstable region, while unshaded  region represents stable region by observing evolution of $m=0(l=0)$ scalar mode.
Rotating scalarized black holes are possible to emerge in the shaded region. The starting point (left) of threshold curve is at [$\log(0.107)$, 0.998] and the ending point is at [$\log(2030.5)$, 0.00025].}
\end{figure*}

Now, we compute the late-time tails of  a perturbed scalar numerically in time-domain by employing the (2+1)-dimensional hyperboloidal foliation method~\cite{Gao:2018acg,Doneva:2020nbb,Zhang:2020pko}. Here, we do not wish  to describe the details of computation steps because we have recovered the main results (especially Fig. 6) in~\cite{Gao:2018acg} exactly.
The Kerr black hole is unstable under $m=0(l=0)$ scalar mode  in the certain region of the parameter space $(\log[\alpha],a)$, as is shown in Fig. 2.
Our picture shows a bigger unstable region because we include the region for $\alpha\in [0.107, 2030.5]$.
The curve represents the threshold one which is the boundary between stable and unstable regions. For the points in the region above the curve (shaded region),
there exist unstable scalar modes and thus, the Kerr black hole is unstable. This region denotes onset of tachyonic instability, implying emergence of the rotating scalarized black holes. Clearly,  the shaded region in Fig. 2  represents an $\alpha$-dependent bound for tachyonic instability.
For the points in the region below the curve (unshaded region),
there exist stable scalar modes and thus, the Kerr black hole is stable. This region has nothing to do with onset of tachyonic instability.
The shaded region is compared to that of EGBS theory~\cite{Cunha:2019dwb} in the sense that the latter includes the tachyonic instability line on the $\log[\alpha]$-axis for  static black holes in the non-rotating limit of $a\to 0$.
In addition, it is hard to realize the unstable region in the extremal  limit of $a\to 1$ for small $\alpha$. This means that the highest rotation excludes
the spontaneous scalarization.

\subsection{$\alpha<0$ case}
Recently, it was shown that the tacyonic instability condition  of Kerr black holes is found as a bound of $a/M \ge 0.5$ in the EGBS theory with  $\alpha<0$ by solving the (1+1)-dimensional linearized scalar equation~\cite{Dima:2020yac}. This was confirmed analytically in~\cite{Hod:2020jjy} and numerically in~\cite{Doneva:2020nbb}. This is so-called the spin-induced black hole scalarization in the EGBS theory with negative coupling constant   because this never occurs in the static black holes.
Hence, it is very curious to see whether such a bound occurs or not in the ECSS theory  with negative coupling constant.
First of all, we employ  the (2+1)-dimensional hyperboloidal foliation method to find the $m=0(l=0)$  time-evolving scalar mode.
The result displays in Fig. 3 which is the same figure as in Fig. 2. The shaded region in Fig. 3  represents an $\alpha$-dependent bound for tachyonic instability.
Importantly, this implies that there is no $\alpha$-independent bound for the rotation parameter $a$  to describe the onset of tachyonic instability when considering $-\alpha\in[0.107,2030.5]$.
This result could be confirmed by checking ``no $\alpha$-independent bound" analytically in the (1+1)-dimensional linearized scalar equation.
\begin{figure*}[t!]
   \centering
  \includegraphics{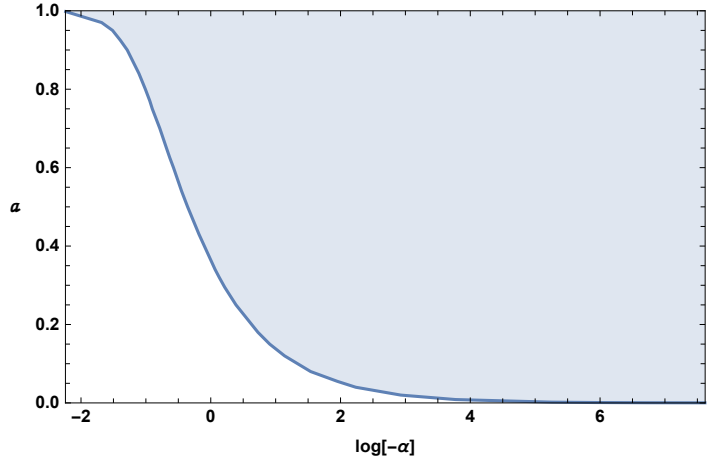}
\caption{$\alpha$-dependent instability bound for negative $\alpha$. Shaded region denotes unstable region, while unshaded  region represents stable region by observing evolution of $m=0(l=0)$ scalar mode.
Rotating scalarized black holes are possible to emerge in the shaded region.  The starting point (left) of threshold curve  is at [$\log(0.107)$, 0.998] and the ending point is at [$\log(2030.5)$, 0.00025]. This picture shows that there is no $\alpha$-independent instability bound for $a$.}
\end{figure*}
For this purpose, let us introduce a projection of Eq.(\ref{phi-eq2}) onto a basis of spherical harmonics $Y_{lm}(\theta,\varphi)$ to obtain the coupled (1+1)-dimensional evolution equation~\cite{Dima:2020yac}
\begin{eqnarray}\label{perteq}
&[(r^2+a^2)^2-a^2\Delta(1- c^m_{ll})]\ddot{\psi}_{l}\nonumber\\
&+a^2\Delta(c_{l,l+2}^m\ddot{\psi}_{l+2}+c_{l,l-2}^m\ddot{\psi}_{l-2})+4iamMr\dot{\psi}_l \nonumber\\
&-(r^2+a^2)^2\psi''_l-\Big[2iam(r^2+a^2)-2a^2\frac{\Delta}{r}\Big]\psi'_l\nonumber\\
&+\Delta \left[l(l+1)+\frac{2M}{r}-\frac{a^2}{r^2}+\frac{2iam}{r} \right]\psi_l\nonumber\\
&+\Delta\sum_{j}< l m | \mu^2_{\rm eff}[r^2+(a\cos\theta)^2]| j m>\psi_j=0 \,,\\
&\psi_{l}(t,r)\equiv \int   r \delta \phi Y_{l0}Y^*_{l0}d\Omega\equiv < l m=0 | r \delta  \phi  | l m=0> \,,\label{psi-eq}\\
\label{coscoup}
&c^m_{jl}\equiv < l m | \cos^2\!\theta | j m> \nonumber\\&= \frac{\delta_{lj}}{3}+\frac{2}{3}\sqrt{\frac{2j+1}{2l+1}}\langle j,2,m,0|l,m\rangle \cdot \langle j,2,0,0|l,0\rangle\,.
\end{eqnarray}
Here  $< j_1,j_2,m_1,m_2|j_3,m_3>$ are the Clebsch-Gordan coefficients.
We note that the evolution modes of different $m$ decouples because of the  axisymmetry.
Also, even $l$ and odd $l$ modes decouple because of the reflection symmetry.
According to the Hod's approach with $m=0$ mode~\cite{Hod:2020jjy}, at the onset of tachyonic instability, the mixed term of
$\Delta\sum_{j}< l m | \mu^2_{\rm eff}[r^2+(a\cos\theta)^2]| j m>\psi_j$ can be replaced by a single term as
\begin{equation}
\Delta< l_1 m=0 | \mu^2_{\rm eff}[r^2+(a\cos\theta)^2]| l_2 m=0>\psi_{l_1}
\end{equation}
at asymptotically late times.

Importantly, the spontaneous scalarization is related to  an effective  binding potential well in the vicinity of $r=r_+$  whose two turning points of $r_{\rm in}$ and $r_{\rm out}$ are classified by the relation $r_{\rm out}\ge r_{\rm in}=r_+$. A threshold  critical black hole with $a=a_{\rm crit}$  which marks the boundary between Kerr and rotating scalarized black holes is characterized by  the presence of a degenerate  binding potential well whose two turning points
merge at the outer horizon $r=r_+$ as
\begin{equation}
< l_1 m=0 | \mu^2_{\rm eff}[r^2+(a\cos\theta)^2]| l_2 m=0>|_{r=r_{\rm in}=r_{\rm out}=r_+(a_{\rm crit})}=0,~~ {\rm for}~a=a_{\rm crit}
\end{equation}
in the limit of $\alpha \to -\infty$.
In this case, the critical rotation parameter $a_{\rm crit}$ is determined by the resonance condition of
\begin{equation}
\int_0^\pi \frac{ar_+\cos \theta(3r_+^2-a^2\cos^2\theta)(r_+^2-3a^2\cos^2\theta)}{(r_+^2+a^2\cos^2\theta)^5}Y_{l_10}Y_{l_20} \sin\theta d\theta =0,~~ {\rm for}~a=a_{\rm crit}.\label{res-con}
\end{equation}
In order to solve Eq.(\ref{res-con}) for $a_{\rm crit}$ analytically, one introduces $\hat{a}=a_{\rm crit}/r_+$ and $x=\hat{a} \cos \theta$.
Then, Eq.(\ref{res-con}) takes the form
\begin{equation} \label{con-im}
\int^{\hat{a}}_{-\hat{a}} \frac{x(3-10x^2+3x^4)}{(1+x^2)^5}  Y_{l_10}(x/\hat{a})Y_{l_20}(x/\hat{a})dx=0.
\end{equation}
One finds easily that this integration is zero for all $l_1=l_2=l=0,1,\cdots$ because the integrand is always an odd function ($Y_{l0}^2=[(2l+1)/4\pi]p_l^2\to$ even function).
It seems that  we may  not determine $a_{\rm crit}$ for $l=0,1,\cdots$ modes by using the Hod's approach.
In the asymptotic limit of $l_1=l_2 =l\to \infty$, $Y^2_{l0}\to \delta(\theta)$ around the poles of  $\theta=0,\pi$.
In this case, Eq.(\ref{con-im}) leads to a simple condition for the  resonance
\begin{equation}
\hat{a}(3-10 \hat{a}^2+3\hat{a}^4) =0
\end{equation}
which solves to give $\hat{a}=0,0.577$.
This implies
\begin{equation}
\Big(\frac{a}{M}\Big)_{\rm crit}=0,\quad 0.866
\end{equation}
for  the critical black hole rotation parameter.  The former   confirms ``no $\alpha$-independent bound"  that   appeared in the Fig.3,
while the second  is just  the enhanced (instability) bound [$(a/M)>(a/M)_{\rm crit}=0.866$] for the  scalarization in the ECSS theory with negative coupling parameter.
At this stage, it is not easy to prove the existence of this enhanced bound  by choosing $l\to \infty$-scalar mode in the ECSS theory with negative coupling parameter.

On the other hand, for the EGBS theory with negative coupling paramter~\cite{Hod:2020jjy},  the resonance condition leads to
 \begin{equation} \label{egbs-im}
\int^{\hat{a}}_0 \frac{1-15x^2+15x^4-x^6}{(1+x^2)^5}  Y_{l_10}(x/\hat{a})Y_{l_20}(x/\hat{a})dx=0.
\end{equation}
 One found that  for $l_1=l_2=0$, $(a/M)_{\rm crit}=0.9008$, for $l_1=l_2=1$, $(a/M)_{\rm crit}=0.6409$, and
 for $l_1=l_2=\infty$, $(a/M)_{\rm crit}=0.5$. From the last, one recovers the $\alpha$-independent bound of $(a/M)\ge 0.5$ for tachyonic instability~\cite{Dima:2020yac}.
 This implies that
 slowly rotating Kerr black holes with $(a/M)<0.5$ could not develop the  tachyonic instability and could not have rotating scalarized black holes.

\section{Discussions}
We have investigated the onset of rotating scalarized black holes in the ECSS theory with coupling parameter $\alpha$.
For positive coupling parameter ($\alpha>0$), the Kerr black hole is unstable against  scalar-mode
perturbation.  In this case, we obtained  the $\alpha$-dependent instability bound shown in Fig. 2. 
In  case of $\alpha<0$, it is shown that  the Kerr black hole is unstable (the $\alpha$-dependent instability bound shown in Fig. 3) without any $\alpha$-independent instability bound on the rotation parameter $a$.
The disappearance of the $a$-bound  is  confirmed analytically
as $(a/M)> 0$ by considering $l\to \infty$-scalar mode in the limit of $\alpha\to -\infty$. 
In addition, we found the other bound of $(a/M)> 0.866$ which may be  identified with an instability bound
for scalarization in  the ECSS theory with negative coupling constant. If this bound occurs really, it implies that rotating Kerr black holes with $(a/M)<0.866$ do not develop
the tachyonic instability and could not obtain rotating scalarized black holes. However, we could not prove the existence of this bound directly.

 \vspace{1cm}

{\bf Acknowledgments}
\vspace{1cm}

This work was supported by the National Research Foundation of Korea (NRF) grant funded by the Korea government (MOE)
 (No. NRF-2017R1A2B4002057). We thank Y. Huang for helpful discussions.


\vspace{1cm}





\newpage
\begin{thebibliography}{99}



\bibitem{Doneva:2017bvd}
  D.~D.~Doneva and S.~S.~Yazadjiev,
  Phys.\ Rev.\ Lett.\  {\bf 120}, no. 13, 131103 (2018)
  doi:10.1103/PhysRevLett.120.131103
  [arXiv:1711.01187 [gr-qc]].

\bibitem{Silva:2017uqg}
  H.~O.~Silva, J.~Sakstein, L.~Gualtieri, T.~P.~Sotiriou and E.~Berti,
  Phys.\ Rev.\ Lett.\  {\bf 120}, no. 13, 131104 (2018)
  doi:10.1103/PhysRevLett.120.131104
  [arXiv:1711.02080 [gr-qc]].

\bibitem{Antoniou:2017acq}
  G.~Antoniou, A.~Bakopoulos and P.~Kanti,
  Phys.\ Rev.\ Lett.\  {\bf 120}, no. 13, 131102 (2018)
  doi:10.1103/PhysRevLett.120.131102
  [arXiv:1711.03390 [hep-th]].


\bibitem{Herdeiro:2018wub}
  C.~A.~R.~Herdeiro, E.~Radu, N.~Sanchis-Gual and J.~A.~Font,
  Phys.\ Rev.\ Lett.\  {\bf 121}, no. 10, 101102 (2018)
  doi:10.1103/PhysRevLett.121.101102
  [arXiv:1806.05190 [gr-qc]].


\bibitem{Myung:2018iyq}
  Y.~S.~Myung and D.~C.~Zou,
  Phys.\ Rev.\ D {\bf 98}, no. 2, 024030 (2018)
  doi:10.1103/PhysRevD.98.024030
  [arXiv:1805.05023 [gr-qc]].


\bibitem{Myung:2018vug}
  Y.~S.~Myung and D.~C.~Zou,
  Eur.\ Phys.\ J.\ C {\bf 79}, no. 3, 273 (2019)
  doi:10.1140/epjc/s10052-019-6792-6
  [arXiv:1808.02609 [gr-qc]].


\bibitem{Zou:2020zxq}
  D.~C.~Zou and Y.~S.~Myung,
  Phys.\ Rev.\ D {\bf 102}, no. 6, 064011 (2020)
  doi:10.1103/PhysRevD.102.064011
  [arXiv:2005.06677 [gr-qc]].

\bibitem{Ayzenberg:2013wua}
  D.~Ayzenberg, K.~Yagi and N.~Yunes,
  Phys.\ Rev.\ D {\bf 89}, no. 4, 044023 (2014)
  doi:10.1103/PhysRevD.89.044023
  [arXiv:1310.6392 [gr-qc]].


\bibitem{Myung:2019wvb}
  Y.~S.~Myung and D.~C.~Zou,
  Int.\ J.\ Mod.\ Phys.\ D {\bf 28}, no. 09, 1950114 (2019)
  doi:10.1142/S0218271819501141
  [arXiv:1903.08312 [gr-qc]].


\bibitem{Molina:2010fb}
  C.~Molina, P.~Pani, V.~Cardoso and L.~Gualtieri,
  Phys.\ Rev.\ D {\bf 81}, 124021 (2010)
  doi:10.1103/PhysRevD.81.124021
  [arXiv:1004.4007 [gr-qc]].

\bibitem{Moon:2011fw}
  T.~Moon and Y.~S.~Myung,
  Phys.\ Rev.\ D {\bf 84}, 104029 (2011)
  doi:10.1103/PhysRevD.84.104029
  [arXiv:1109.2719 [gr-qc]].

\bibitem{Kimura:2018nxk}
  M.~Kimura,
  Phys.\ Rev.\ D {\bf 98}, no. 2, 024048 (2018)
  doi:10.1103/PhysRevD.98.024048
  [arXiv:1807.05029 [gr-qc]].

\bibitem{Macedo:2018txb}
  C.~F.~B.~Macedo,
  Phys.\ Rev.\ D {\bf 98}, no. 8, 084054 (2018)
  doi:10.1103/PhysRevD.98.084054
  [arXiv:1809.08691 [gr-qc]].

\bibitem{Cunha:2019dwb}
  P.~V.~P.~Cunha, C.~A.~R.~Herdeiro and E.~Radu,
  Phys.\ Rev.\ Lett.\  {\bf 123}, no. 1, 011101 (2019)
  doi:10.1103/PhysRevLett.123.011101
  [arXiv:1904.09997 [gr-qc]].

\bibitem{Collodel:2019kkx}
  L.~G.~Collodel, B.~Kleihaus, J.~Kunz and E.~Berti,
  Class.\ Quant.\ Grav.\  {\bf 37}, no. 7, 075018 (2020)
  doi:10.1088/1361-6382/ab74f9
  [arXiv:1912.05382 [gr-qc]].

\bibitem{Dima:2020yac}
  A.~Dima, E.~Barausse, N.~Franchini and T.~P.~Sotiriou,
  arXiv:2006.03095 [gr-qc].

\bibitem{Hod:2020jjy}
  S.~Hod,
  Phys.\ Rev.\ D {\bf 102}, no. 8, 084060 (2020)
  doi:10.1103/PhysRevD.102.084060
  [arXiv:2006.09399 [gr-qc]].

\bibitem{Doneva:2020nbb}
  D.~D.~Doneva, L.~G.~Collodel, C.~J.~Krüger and S.~S.~Yazadjiev,
  Phys.\ Rev.\ D {\bf 102}, no. 10, 104027 (2020)
  doi:10.1103/PhysRevD.102.104027
  [arXiv:2008.07391 [gr-qc]].


\bibitem{Herdeiro:2020wei}
  C.~A.~R.~Herdeiro, E.~Radu, H.~O.~Silva, T.~P.~Sotiriou and N.~Yunes,
  arXiv:2009.03904 [gr-qc].



\bibitem{Berti:2020kgk}
  E.~Berti, L.~G.~Collodel, B.~Kleihaus and J.~Kunz,
  arXiv:2009.03905 [gr-qc].




\bibitem{Zhang:2020pko}
  S.~J.~Zhang, B.~Wang, A.~Wang and J.~F.~Saavedra,
  arXiv:2010.05092 [gr-qc].




\bibitem{Gao:2018acg}
  Y.~X.~Gao, Y.~Huang and D.~J.~Liu,
  Phys.\ Rev.\ D {\bf 99}, no. 4, 044020 (2019)
  doi:10.1103/PhysRevD.99.044020
  [arXiv:1808.01433 [gr-qc]].




\end{thebibliography}
\end{document}